\documentclass[showpacs,aps,twocolumn]{revtex4}
\usepackage{epsf}
\newcommand{\mb}[1]{ { \mbox{\boldmath{$#1$}}}  } 
\begin{document}

%------------------------------------------------------------------- 
\title{Replicas of the Fano resonances induced by phonons in 
        a subgap Andreev tunneling}

     \author{J.\ Bara\'nski and T.\ Doma\'nski}
\affiliation{
             Institute of Physics, 
	     M.\ Curie Sk\l odowska University, 
             20-031 Lublin, Poland} 
      \date{\today}
%-------------------------------------------------------------------

\begin{abstract}
We study influence of the phonon modes on a subgap spectrum and Andreev 
conductance for the double quantum dot vertically coupled between a metallic 
and superconducting lead. For the monochromatic phonon reservoir we obtain 
replicas of the interferometric Fano-type structures appearing simultaneously
in the particle and hole channels. We furthermore confront the induced on-dot 
pairing with the electron correlations and investigate how the phonon modes 
affect the zero-bias signature of the Kondo effect in Andreev conductance. 
\end{abstract}  


\maketitle

\section{Introduction} 

Electron transport through nano-size transistors containing the quantum dots, 
molecules and/or nanowires is determined by the available energy levels (tunable 
by external gate voltage) and strongly depends on the Coulomb interactions 
\cite{corr_theor}. Discretization of the energy levels is responsible for 
oscillations of the differential conductance upon varying the gate voltage, 
whereas the correlation effects lead to the Coulomb blockade and can induce 
(at low temperatures) the Kondo resonance enhancing the zero-bias conductance 
to a unitary value $\frac{2e^{2}}{h}$ \cite{GoldhaberGordon-98,corr_exper}. Besides 
promising perspectives for the applications in modern electronics/spintronics 
the nanoscopic structures represent also valuable testing grounds for probing 
the many-body effects. Magnetic, superconducting or other types of orderings 
absorbed from the external leads can be confronted with the on-dot electron 
correlations in a fully controllable manner. 

In this regard, especially interesting are the heterojunctions where the quantum 
dots (QDs) are in contact with the superconducting (S) electrodes. Nonequilibrium 
charge transport can occur there either via the usual single particle tunneling 
(upon breaking the electron pairs) or by activating the anomalous (Andreev 
or Josephson) channels. The resulting currents are sensitive to 
a competition between the induced on-dot pairing and the Coulomb repulsion. 
In such context there have been experimentally explored the signatures of 
$\pi$-junction \cite{lit_7}, Josephson effect \cite{lit_8}, superconducting quantum 
interference \cite{lit_9}, quantum entanglement by splitting the Copper pairs 
\cite{lit_10}, multiple Andreev scattering \cite{lit_11}, and interplay of the 
on-dot pairing with the Kondo effect \cite{lit_12,Deacon-10,Franceschi-12}.
These and similar related activities have been discussed theoretically 
by various groups \cite{Yamada-11,Rodero-11,interplay,Droste-12,Konig-12}.

\begin{figure}
\epsfxsize=9cm\centerline{\epsffile{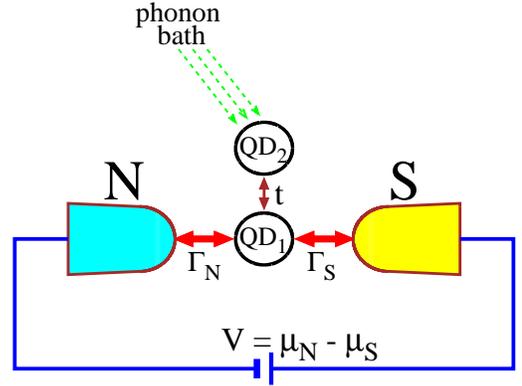}}
\vspace{-0.7cm}
\caption{(color online) Schematic view of the double quantum dot coupled 
in T-shape configuration between the metallic (N) and  superconducting 
(S) electrodes where an external phonon bath affects the side-attached 
quantum dot (QD2).}
\label{scheme}
\end{figure}

Since in practical realizations the nanoscopic objects are never entirely separated 
from an environment (e.g.\ a given substrate or external photon/phonon quanta) 
therefore transport properties are also affected by the interference effects. 
A convenient prototype for studying such phenomena is the tunneling setup
shown in figure \ref{scheme}, where the central quantum dot is coupled to 
the side-attached quantum dot and eventually to other degrees of freedom.
Transport through this T-shape double quantum dot (DQD) occurs predominantly 
via the central quantum dot (QD$_{1}$), whereas electron leakage to/from 
the side-coupled quantum dot (QD$_{2}$) brings in the interference effects. 
In the case of both normal (N) electrodes and assuming a weak interdot 
coupling it has been argued \cite{Maruyama-04,Zitko-10} that the
differential conductance should reveal the asymmetric Fano-type 
lineshapes. This fact has been indeed observed experimentally \cite{Sasaki-09}.
 
Recently we have explored the interferometric patterns for the DQD case 
placed between the metallic and  superconducting electrodes \cite{Baranski-11}. 
In such hetero\-structures  the interferometric lineshapes appear simultaneously 
at negative and positive energies because of the mixed particle and hole 
degrees in the effective quantum dot spectrum \cite{Balatsky-06}. These 
effects manifest themselves in the subgap Andreev conductance \cite{Deacon-10}. 
Stability of interferometric Fano structures on a dephasing by 
external fermionic bath have been analyzed in Refs \cite{Baranski-12,Michalek-12}. 
Here we extend such study addressing the role of bosonic bath 
in the setup displayed in Fig.\ \ref{scheme}. We argue that monochromatic 
phonon bath induces a number of the Fano-type replicas depending on 
the adiabadicity ratio $\lambda/\omega_{0}$. 

In the next section we introduce the microscopic model and briefly specify  
characteristic energy scales. We also discuss the formal aspects concerning
the adopted approximations. In the section III we analyze spectroscopic 
fingerprints of the phonon modes for the case of uncorrected quantum dots. 
Finally, in section IV, we address the correlation effects (Coulomb 
blockade and Kondo physics) due to the on-dot repulsion between 
the opposite spin electrons. Appendix A provides phenomenological 
arguments for the Fano-type interferometric patters of the double quantum 
dot structures.

\section{Microscopic model}

The double quantum dot heterojunction shown in Fig.\ \ref{scheme} 
can be described by the Anderson-type Hamiltonian
\begin{eqnarray} 
\hat{H} =   \hat{H}_{leads} + \hat{H}_{DQD} + \hat{H}_{T}
\label{model} 
\end{eqnarray}
where $\hat{H}_{leads}\!=\!\hat{H}_{N}\!+\!\hat{H}_{S}$ denote the normal 
$N$ and superconducting $S$ charge reservoirs, $\hat{H}_{DQD}$ refers 
to both quantum dots (together with the phonon bath) and the last 
term $\hat{H}_{T}$ describes the hybridization to external leads. We 
treat the conducting lead as a Fermi gas $\hat{H}_{N}=\sum_{{\bf k},\sigma} 
\xi_{{\bf k}N} \hat{c}_{{\bf k} \sigma N}^{\dagger} \hat{c}_{{\bf k}\sigma 
N}$ and we assume that isotropic superconductor is described by the BCS 
Hamiltonian $\hat{H}_{S} \!=\!\sum_{{\bf k},\sigma}  \xi_{{\bf k}S} 
\hat{c}_{{\bf k} \sigma S }^{\dagger}  \hat{c}_{{\bf k} \sigma S} 
\!-\! \Delta \sum_{\bf k} ( \hat{c}_{{\bf k} \uparrow S }^{\dagger} 
\hat{c}_{-{\bf k} \downarrow S }^{\dagger} \!+\! \hat{c}_{-{\bf k} 
\downarrow S} \hat{c}_{{\bf k} \uparrow S})$. The operators $\hat{c}
_{{\bf k} \sigma \beta} ^{({\dagger})}$ correspond to annihilation 
(creation) of the itinerant electrons with spin $\sigma=\uparrow,
\downarrow$ and energies $\xi_{{\bf k}\beta}\!=\!\varepsilon_{{\bf k}\beta} 
\!-\!\mu_{\beta}$ are measured with respect to 
the chemical potentials $\mu_{\beta}$.

The double quantum dot along with the phonon bath is described by 
following local part
\begin{eqnarray} 
&& \hat{H}_{DQD} = \sum_{\sigma,i} \varepsilon_{i}  
\hat{d}^{\dagger}_{i \sigma} \hat{d}_{i \sigma} +
t \sum_{\sigma} \left(  \hat{d}_{1\sigma}^{\dagger}  
\hat{d}_{2\sigma} \!+\! \mbox{\rm H.c.} \right)
\\ & & + \sum_{i}U_{i} \; 
\hat{d}^{\dagger}_{i \uparrow} \hat{d}_{i \uparrow} \; 
\hat{d}^{\dagger}_{i \downarrow} \hat{d}_{i \downarrow} 
+\omega_0 \hat{a}^{\dagger}\hat{a}+\lambda \sum_{\sigma}\hat{d}
^{\dagger}_{2 \sigma} \hat{d}_{2 \sigma}(\hat{a}^{\dagger}+\hat{a}) ,
\label{DQD} 
\nonumber
\end{eqnarray} 
where we use the standard notation for the annihilation (creation) 
operators $\hat{d}_{i}^{({\dagger})}$ for electrons at each quantum 
dot QD$_{i=1,2}$. Their energy levels are denoted by $\varepsilon_{i}$ 
and $U_{i}$ refer to the on-dot Coulomb potentials. Since we are
interested in the Fano-type interference we focus on the electron
transport only via the central quantum dot 
\begin{eqnarray} 
\hat{H}_{T} &=& \sum_{\beta = N,S} \sum_{{\bf k},\sigma} 
\left( V_{{\bf k} \beta} \; \hat{d}_{1 \sigma}^{\dagger}  
\hat{c}_{{\bf k} \sigma \beta } + \mbox{\rm H.c.} \right) .
\label{hybr}
\end{eqnarray} 
This situation can be extended to more general cases when electron
tunneling directly involves both the quantum dots. For clarity reasons
we postpone such analysis for the future studies.

\subsection{Outline of the formalism}

Energy spectrum and transport properties of the system (\ref{model}) can
be inferred from the matrix Green's function $\mb{G}_i(\tau_1 , \tau_2)=
-i\hat{T_\tau}\langle \hat{\Psi}_i(\tau_1) \hat{\Psi}^{\dagger}_i(\tau_2)$ 
defined in a representation of the Nambu spinors $\hat{\Psi}^{\dagger}_i
\equiv(\hat{d}^{\dagger}_{i \uparrow}, \hat{d}_{i \downarrow})$, $\hat{\Psi}
_i\equiv(\hat{\Psi}^{\dagger}_i)^{\dagger}$. In the equilibrium conditions 
(for $\mu_{N}\!=\mu_{S}$) such matrix Green's function depends only on time 
difference $\tau_1\!-\!\tau_2$. The corresponding Fourier transform 
can be then expressed by the Dyson equation
\begin{eqnarray}
\mb{G}_i^{-1}(\omega)=\mb{g}_i^{-1}(\omega)-\mb{\Sigma}_i^{0}(\omega)
-\mb{\Sigma}_i^{U}(\omega) ,
\label{Dyson}
\end{eqnarray}
where the bare propagators $\mb{g}_{i}(\omega)$ of uncorrelated quantum dots
are given by
\begin{eqnarray}
\mb{g}_i^{-1}(\omega)=\left(\begin{array}{cc}\omega-\varepsilon_i &0
\\0& \omega+\varepsilon_i \end{array}\right) .
\end{eqnarray}
First part of the selfenergy $\mb{\Sigma}_i^{0}(\omega)$ comes from a combined 
effect of the interdot coupling, hybridization of the central QD$_{1}$ with 
the external leads (\ref{hybr}) and the phonon bath contribution acting on QD$_{2}$. 
The other term $\mb{\Sigma}_i^{U}(\omega)$ appearing in (\ref{Dyson}) accounts 
for the many-body effects originating from the on-dot Coulomb repulsion $U_{i}$.

Let us begin by first specifying the selfenergy $\mb{\Sigma}_1^{0}(\omega)$  
for the uncorrelated central quantum dot. The usual diagrammatic approach 
yields
\begin{eqnarray}
\mb{\Sigma}_1^{0}(\omega)=\sum_{{\bf k}, \beta} V_{{\bf k} \beta} 
\mb{g}_{\beta}({\bf k}, \omega) \; V_{{\bf k} \beta}^{*} + 
t \; \mb{G}_2(\omega)\; t^{*} ,
\label{Sigma_0_QD1}
\end{eqnarray}
where $g_{\beta}({\bf k}, \omega)$ denote the matrix Green's functions 
of the leads. In particular, we have  for the normal lead 
\begin{eqnarray} 
{\mb g}_{N}({\bf k}, \omega) = 
\left( \begin{array}{cc}  
\frac{1}{\omega-\xi_{{\bf k}N}} & 0 \\ 
0 &  
\frac{1}{\omega+\xi_{{\bf k}N}}
\end{array}\right)
\label{gN}
\end{eqnarray} 
and for the superconducting electrode 
\begin{eqnarray} 
{\mb g}_{S}({\bf k}, \omega) = 
\left( \begin{array}{cc}  
\frac{u^{2}_{\bf k}}{\omega-E_{\bf k}}+\frac{v^{2}_{\bf k}}
{\omega+E_{\bf k}} \hspace{0.2cm} & \frac{-u_{\bf k}v_{\bf k}}
{\omega-E_{\bf k}}+\frac{u_{\bf k}v_{\bf k}}{\omega+E_{\bf k}}
\\ 
\frac{-u_{\bf k}v_{\bf k}}{\omega-E_{\bf k}}+
\frac{u_{\bf k}v_{\bf k}}{\omega+E_{\bf k}}
& \frac{u^{2}_{\bf k}}{\omega+E_{\bf k}}+
\frac{v^{2}_{\bf k}}{\omega-E_{\bf k}}
\end{array}\right) 
\label{gS}\end{eqnarray} 
with quasiparticle energy $E_{\bf k}\!=\!\sqrt{\xi_{{\bf k}S}^{2}
+\Delta^{2}}$ and the BCS coefficients 
$u^{2}_{\bf k},v^{2}_{\bf k} = \frac{1}{2} \left[ 1 \pm 
\frac{\xi_{{\bf k}S}}{E_{\bf k}} \right]$,
$u_{\bf k}v_{\bf k} = \frac{\Delta}{2E_{\bf k}}$.
In the wide band limit approximation we assume the constant 
hybridization couplings $\Gamma_{\beta}=2\pi \sum_{\bf k} |V_{{\bf k}
\beta}|^2 \; \delta(\omega \!-\! \xi_{{\bf k}\beta})$ and treat $\Gamma_{N}$
as a convenient unit for energies. We then formally have 
\begin{eqnarray}
\sum_{{\bf k}} |V_{{\bf k}N}|^{2} \; {\mb g}_{\beta}({\bf k},\omega) 
 &=&  -i \frac{\Gamma_{N}}{2} \; 
\left( \begin{array}{cc}  
1 & 0 \\ 0 & 1 \end{array} \right)
\label{Sigma_N}\\
\sum_{{\bf k}} |V_{{\bf k}S}|^{2} \; {\mb g}_{S}({\bf k},\omega) 
&=& -i \frac{\Gamma_{S}}{2} \gamma(\omega)
\left( \begin{array}{cc}  
1 & \frac{\Delta}{\omega} \\ 
 \frac{\Delta}{\omega}  & 1 
\end{array} \right)
\label{Sigma_S}
\end{eqnarray} 
where \cite{Yamada-11} 
\begin{eqnarray}
\gamma(\omega) = \left\{
\begin{array}{ll} 
\frac{|\omega|}{\sqrt{\omega^{2}-\Delta^{2}}}
& \mbox{\rm for }  |\omega| > \Delta , \\
\frac{\omega}{i\sqrt{\Delta^{2}-\omega^{2}}}
& \mbox{\rm for }  |\omega| < \Delta .
\end{array} \right.
\label{gamma}
\end{eqnarray} 
Deep in a subgap regime (i.e.\ for $| \omega| \ll \Delta $) only 
the off-diagonal terms of (\ref{Sigma_S}) survive, approaching
the static value $-\Gamma_{S}/2$. From the physical point of view 
a magnitude $\left| -\Gamma_{S}/2 \right|\equiv \Delta_{d1}$ 
can be interpreted as on-dot pairing gap induced in the QD$_{1}$.
Such situation has been studied in the literature by a number 
of authors \cite{sc_atomic_limit} applying various methods to 
account for the correlation effects $\Sigma_{1}^{U}(\omega)$.

\subsection{Influence of the phonon modes} 

Generally speaking, the phonon modes have both the quantitative and qualitative 
influence on the electron transport through nanodevices \cite{Fransson_book}. 
In particular they can be responsible for such effects as: appearance of
the multiple side-peaks, polaronic shift in the energy levels, lowering 
of the on-dot potential $U_{i}$ (even to the negative values promoting 
the pair hopping \cite{negative-U}), 
suppression of the hybridization couplings $\Gamma_{\beta}$ and 
often serve as a source of the decoherence. These and related 
subjects have been so far studied by many groups, mainly considering the single 
quantum dots coupled to the normal leads \cite{phonons_studies}. 
Here we would like to focus on the different situation (Fig.\ \ref{scheme})
considering the phonon modes coupled to the side-attached quantum dot. 
This is reminiscent of the setup discussed by M.\ B\"uttiker 
\cite{Buttiker-88} except that the fermion reservoir is here 
replaced by the phonon bath. 

In analogy to (\ref{Sigma_0_QD1}) we express the selfenergy 
${\mb{\Sigma}_2^{0}(\omega)}$ of  QD$_{2}$ by the following contributions
\begin{eqnarray}
{\mb{\Sigma}_2^{0}(\omega)} = t\mb{G}_1(\omega)t^{*}
+ \mb{\Sigma}^{ph}_2(\omega) ,
\label{Sigma0_2}
\end{eqnarray}
where $t\mb{G}_1(\omega)t^{*}$ originates from the interdot 
hybridization and the second term is induced by the phonon reservoir. 
Since QD$_{2}$ is assumed to be weakly coupled with the central dot 
we approximate the selfenergy $\mb{\Sigma}^{ph}_2(\omega)$ adopting 
the local solution. The selfenergy $\mb{\Sigma}^{ph}_2(\omega)$ 
can be determined by means of the Lang-Firsov canonical transformation 
%\cite{Lang_Firsov} 
which effectively gives \cite{Mahan_book} 
\begin{eqnarray}
\frac{1}{ \omega - \varepsilon_{2} - \mb{\Sigma}^{ph}_2
(\omega) } = \sum_{l} 
\left(\begin{array}{cc}\frac{{\cal{A}}({l})}{\omega-
\tilde{\varepsilon}_2(l) }&0\\0&\frac{{\cal{A}}({l})}
{\omega+\tilde{\varepsilon}_2(l)}\end{array}\right)
 \label{phonons} 
\end{eqnarray}
with the quasiparticle energies $\tilde{\varepsilon}_{2}(l)
=\varepsilon_{2}-\Delta_{p}+l\omega_{0}$, the corresponding 
polaronic shift $\Delta_p=\lambda^{2}/\omega_0$ and temperature dependent 
spectral weights ${\cal{A}}({l})$  \cite{Fransson_book,Mahan_book}
\begin{eqnarray}
{{\cal{A}}({l})} &=& e^{-g\sqrt{1+2N_{ph}(l)}} \; 
e^{n\omega_{0}/k_{B}T} \label{weight_at_T} \\
& \times & I_{l} \left( 2g^{2} 
\sqrt{N_{ph}(l) \left[ 1 + N_{ph}(l) \right]}\right) .
\nonumber
\end{eqnarray}
As usually, we introduce a dimensionless adiabadicity parameter
$g=\frac{\lambda^2}{\omega_0^{2}}$, $N_{ph}(l)=\left[ e^{\omega_{0}/k_{B}T} 
- 1 \right]^{-1}$ is the Bose-Einstein distribution and $I_{l}$ 
denote the modified Bessel functions. In particular, for the
 ground state the equation (\ref{weight_at_T}) simplifies to
\begin{eqnarray}
\lim_{T \rightarrow 0}{{\cal{A}}({l})}
=e^{-g} \frac{g^l}{l!} \; \theta(l) .
\label{weight_at_0}
\end{eqnarray}
\begin{figure}
\epsfxsize=8cm\centerline{\epsffile{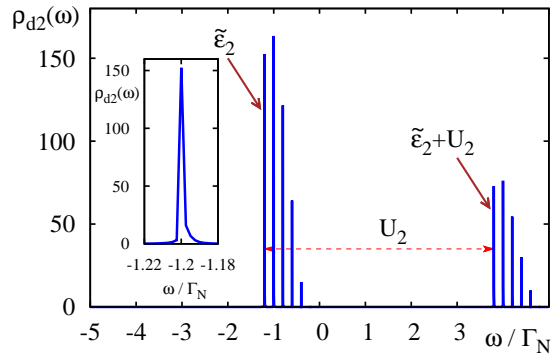}}
\caption{(color online) Spectral function of the side-attached quantum 
dot obtained at low temperature for the  model parameters
$\varepsilon_{2}=-1\Gamma_{N}$, $U_{2}=5\Gamma_{N}$, $\omega_{0}=0.2 
\Gamma_{N}$, $g=1$,  $\varepsilon_{1}=0$, $\Delta = 10 \Gamma_{N}$ 
assuming a weak interdot coupling $t=0.2\Gamma_{N}$.}
\label{dos_QD2}
\end{figure}

In figure \ref{dos_QD2} we illustrate the characteristic spectrum of 
the side-attached quantum dot (see section IV.A for technical details). 
We notice two groups of the narrow peaks. 
The lower one starts from the energy $\tilde{\varepsilon}_{2}=
\varepsilon-\Delta_{p}$ (where $\Delta_{p}=\lambda^{2}/\omega_{0}$ 
is the polaronic shift)  followed by a number of equidistant phonon 
peaks spaced by $\omega_{0}$. The upper phonon branch is separated by 
$U_{2}$ and it manifests the charging effect \cite{Fransson_book}. 
For the coupling $g=1$ we observe only about five phonon peaks but
in the antiadiabatic regime ($g\gg 1$) their number considerably 
increases. Such tendency is shown in section III  discussing
the  spectrum of QD$_{1}$. Let us also stress that the interference peaks 
have a rather tiny but yet finite width $\propto t^{2}/\Gamma_{N}$ \cite{Baranski-11}.

\subsection{Subgap transport}

Charge transport in a subgap regime $|eV|\!<\!\Delta$ is generated 
only by the Andreev mechanism. Electrons coming from the metallic lead 
are then converted into the Cooper pairs in superconductor simultaneously  
reflecting holes back to the normal lead. Such anomalous current $I_{A}(V)$ 
can be expressed by the Landauer-type formula \cite{transport_formula}
\begin{eqnarray} 
I_{A}(V) = \frac{2e}{h} \int d\omega T_{A}(\omega)
\left[ f(\omega\!-\!eV,T)\!-\!f(\omega\!+\!eV,T)\right] ,
\label{I_A}
\end{eqnarray} 
where  $f(\omega,T)=1/\left[ e^{\omega/k_{B}T} + 1 \right]$ is 
the Fermi-Dirac function. Andreev transmittance $T_{A}(\omega)$ 
depends on the off-diagonal part of the Green's function 
${\mb G}_{1}(\omega)$ via  \cite{transport_formula}
\begin{eqnarray} 
T_{A}(\omega) = \Gamma_{N}^{2} \left| G_{1,12}(\omega)\right|^{2} .
\label{Transmittance_A}
\end{eqnarray} 
The transmittance (\ref{Transmittance_A}) can be regarded as 
a qualitative measure of the proximity induced on-dot pairing. 
Under optimal conditions it approaches unity when $\omega$ is close 
to the quasiparticle energies $\pm \sqrt{\varepsilon_{1}^2+
(\Gamma_{S}/2)^{2}}$. The Andreev transmittance $T_{A}(\omega)$ 
is also sensitive to other structures, for instance originating 
from the interferometric effects \cite{Baranski-11,Baranski-12,
Michalek-12}.

In what follwos we shall study the differential Andreev conductance 
\begin{eqnarray} 
G_{A}(V) = \frac{ \partial I_{A}(V)}{\partial V} 
\label{G_A}
\end{eqnarray} 
exploring its dependence on the phonon modes. We start 
this analysis assuming that both quantum dots uncorrelated and 
we next extend it (in section IV) by considering the correlation 
effects. As a general remark, let us notice that the even function 
$T_{A}(-\omega)\!=\!T_{A}(\omega)$ implies the symmetric
Andreev conductance $G_{A}(-V)\!=\!G_{A}(V)$ regardless of any
particular features due to interference, phonons, correlations
or whatever. Physically this is caused by the fact that particle 
and hole degrees of freedom participate equally in the Andreev 
scattering.

\section{Uncorrelated quantum dots}

\begin{figure}
\epsfxsize=8cm\centerline{\epsffile{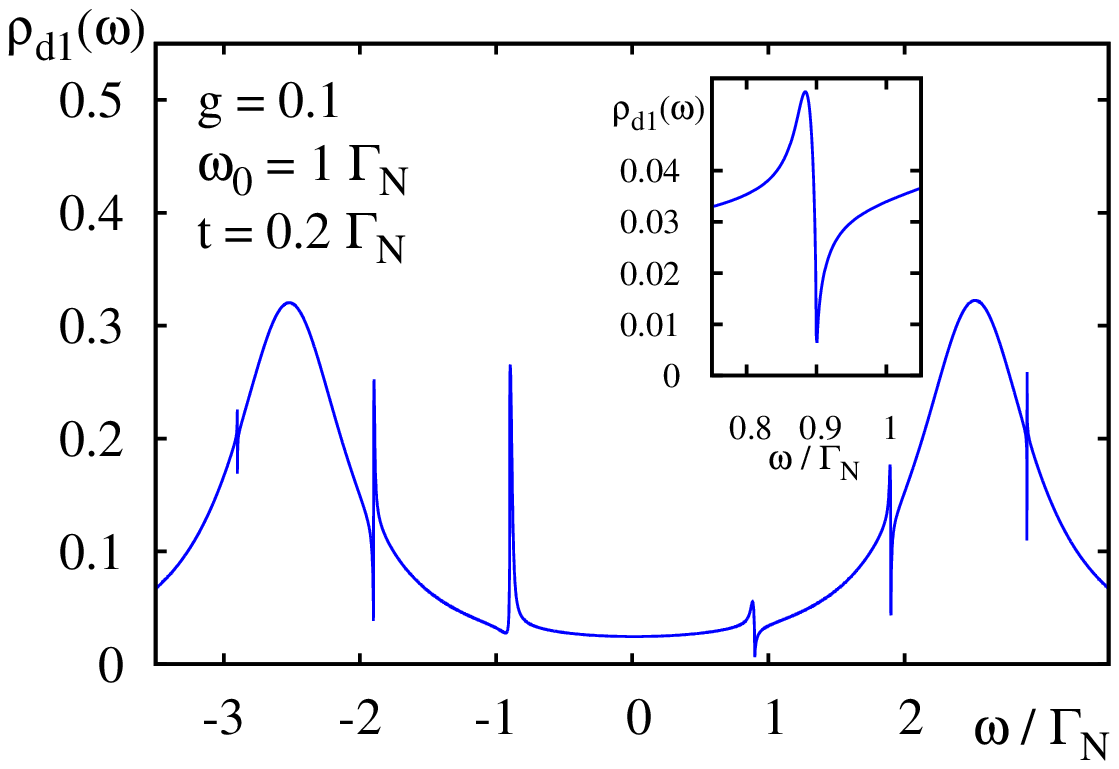}}

\vspace{-0.7cm}
\epsfxsize=8cm\centerline{\epsffile{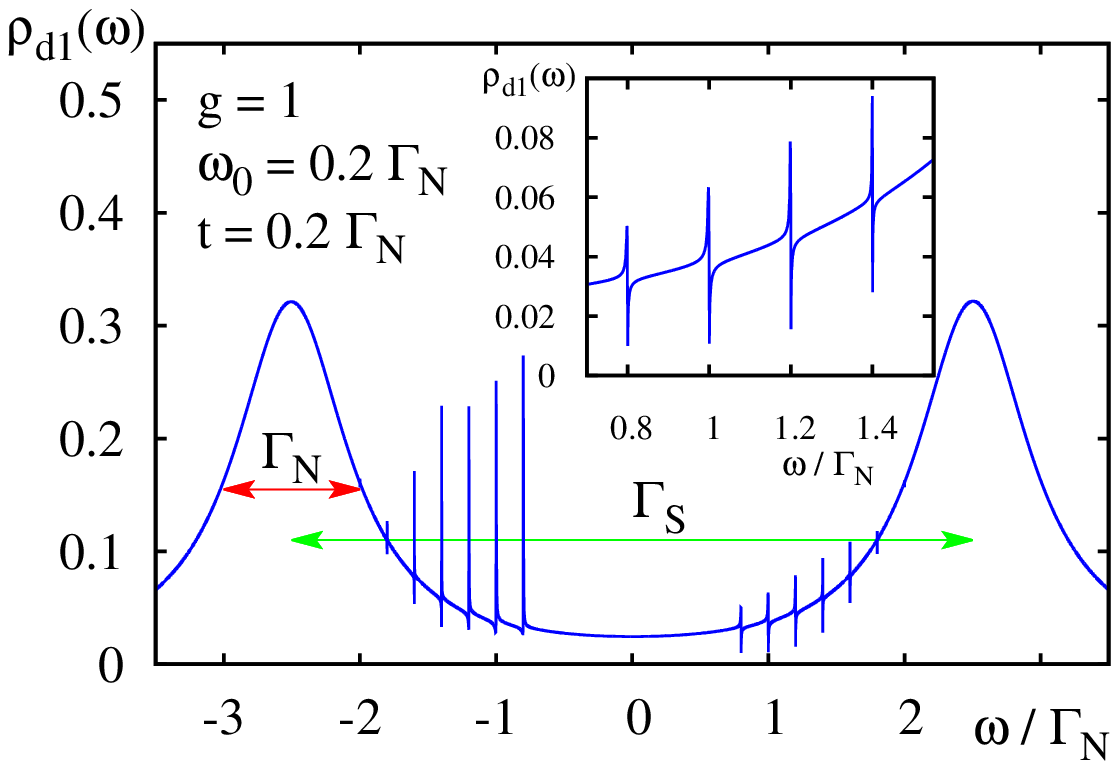}}

\vspace{-0.7cm}
\epsfxsize=8cm\centerline{\epsffile{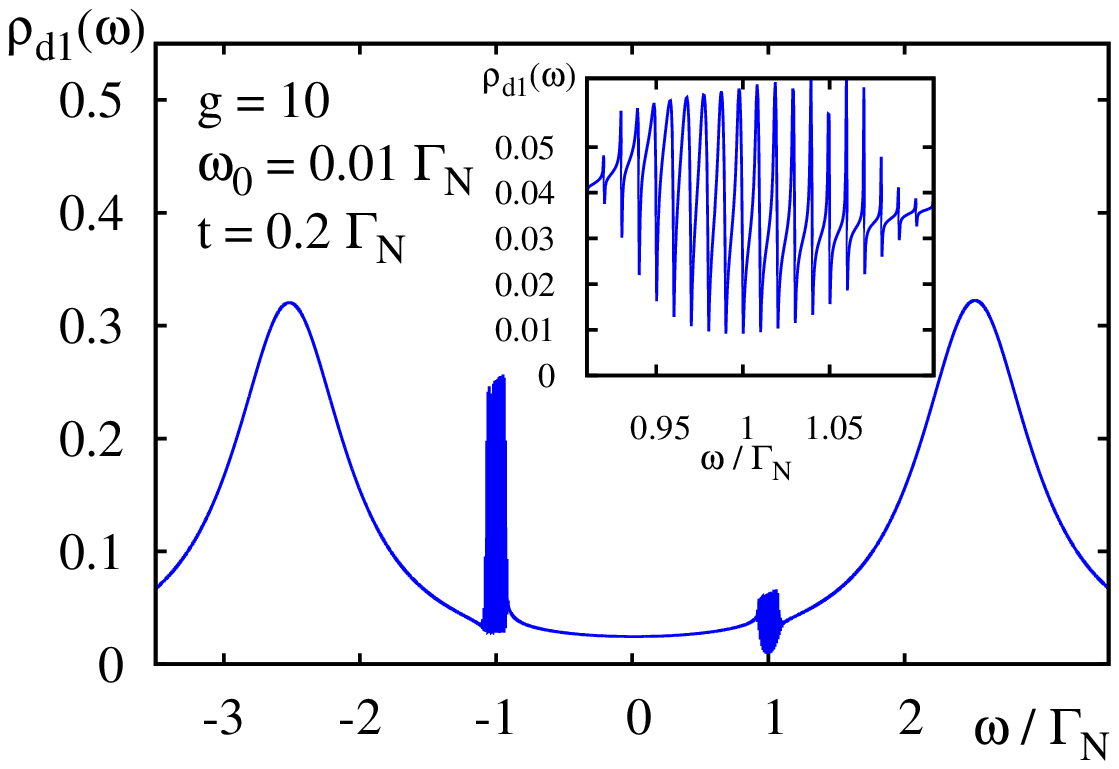}}
\caption{(color online) The interferometric Fano-type lineshapes 
appearing at $T=0$ in the spectral function $\rho_{d1}(\omega)$ of QD$_{1}$.
 Numerical calculations have been done for the uncorrelated quantum 
dots $U_{i}=0$ using the model  parameters $\varepsilon_{1}=0$, 
$\varepsilon_{2}=1\Gamma_{N}$, $t=0.2\Gamma_{N}$, $\Gamma_{S}=5\Gamma_{N}$ 
and $\Delta = 10 \Gamma_{N}$.}
\label{dos_evol}
\end{figure}

Upon neglecting the correlation selfenergies $\Sigma^{U}_{i}
(\omega)\!=\!0$ one has to solve the following coupled equations 
\begin{eqnarray}
\mb{G}_1^{-1}(\omega)&=& \mb{g}^{-1}_1(\omega) - |t|^2 \mb{G}_2(\omega)
 + \frac{1}{2} \left(\begin{array}{cc}i\Gamma_N&\Gamma_S\\ 
\Gamma_S& i\Gamma_N\end{array}\right)\\
\mb{G}_2^{-1}(\omega)&=&  \mb{g}_2^{-1}(\omega)
-|t|^{2} \mb{G}_1(\omega) - \Sigma^{ph}_{2}(\omega) .
\end{eqnarray}
We have computed numerically the matrix Green's functions $\mb{G}_i
(\omega)$ for a mesh of energy points appropriate for the model parameters
(mainly dependent on $\omega_{0}$, $g$ and $t$). Practically already about 
ten iterations proved to yield a fairly convergent solution.

Since the proximity induced on-dot pairing predominantly affects 
the energy region around $\mu_{S}$ we first consider the instructive 
case $\varepsilon_{1}=0$, $\varepsilon_{2}\neq\varepsilon_{1}$. In figure 
\ref{dos_evol} we present the equilibrium spectrum $\rho_{d1}(\omega)$ 
of the central quantum dot obtained for three representative coupling 
constants $g=\lambda/\omega_{0}$ corresponding to the adiabatic limit 
$g\ll 1$ (upper panel), the antiadiabatic regime $g\gg 1$ (bottom 
panel) and the intermediate case (middle panel). On top of two 
Lorentzian peaks centered at the quasiparticle energies 
$\pm \Gamma_{S}/2$ we clearly see formation of the Fano resonances. 
They appear at energies $\tilde{\varepsilon}_{2}+l\omega_{0}$ and 
at their mirror reflections (because of the particle - hole mixing 
\cite{Balatsky-06}). Number of these phonon features depends on 
the adiabadicity parameter $g$. For the adiabatic regime there 
appear only a few phonon features whereas in the opposite antiadiabatic 
limit there is a whole bunch of such narrow structures. In the latter 
case they seem to have an irregular structure, but after closer 
inspection we can clearly see the Fano-type shapes (see the inset).

\begin{figure}
\epsfxsize=7cm\centerline{\epsffile{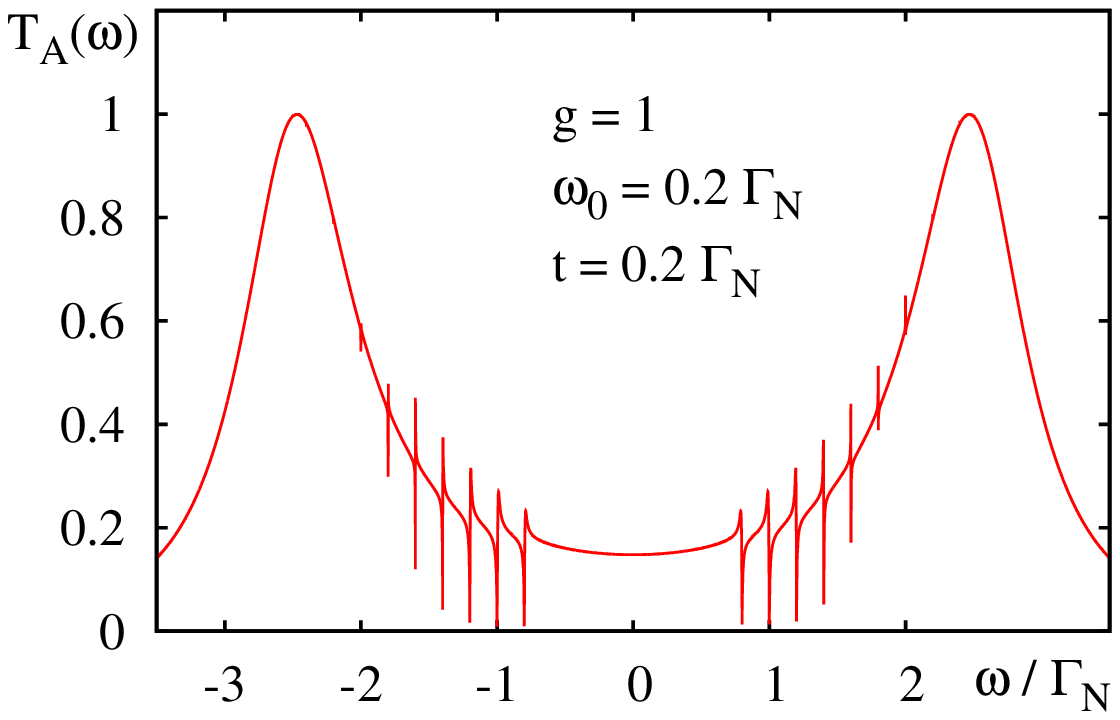}}

\vspace{-0.3cm}
\epsfxsize=7cm\centerline{\epsffile{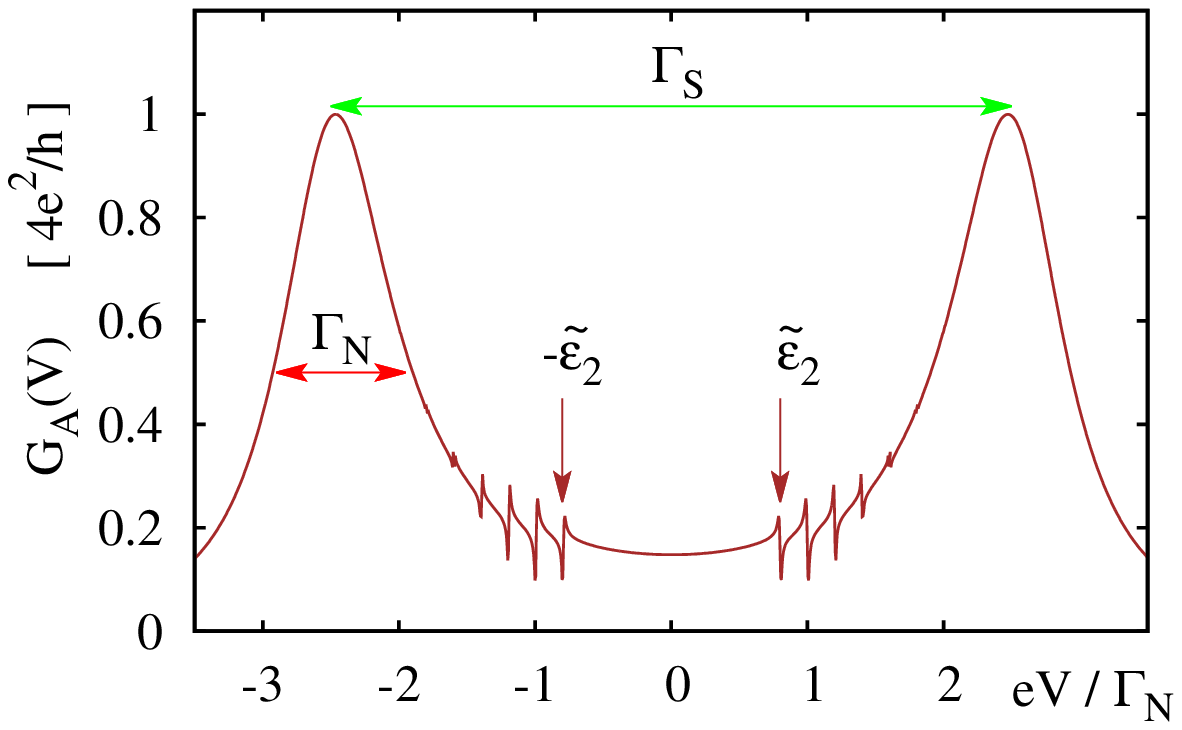}}
\caption{(color online) The Andreev transmittance (upper panel)
and the differential subgap conductance $G_{A}(V)$ (bottom panel)
obtained for the intermediate coupling limit $g=1$ using 
the same parameters as in figure 
\ref{dos_evol}. }
\label{TA_evol}
\end{figure}

Phonon driven replicas of the Fano lineshapes appear also in the 
Andreev transmittance (see Fig.\ \ref{TA_evol}). In a distinction
to the spectral function $\rho_{d1}(\omega)$ the resonances show up
in a symmetrized way due to reasons mentioned in the preceding 
section. Again we notice the broad maxima centered at the subgap
quasiparticle energies $\pm\Gamma_{S}/2$ accompanied by a number
of the Fano-type resonances at $\pm (\tilde{\varepsilon}_{2}+
l\omega_{0}$). Spectroscopic measurements of the Andreev conductance
would thus be able to detect such phonon induced interferometric
features.

\begin{figure}
\epsfxsize=9cm{\epsffile{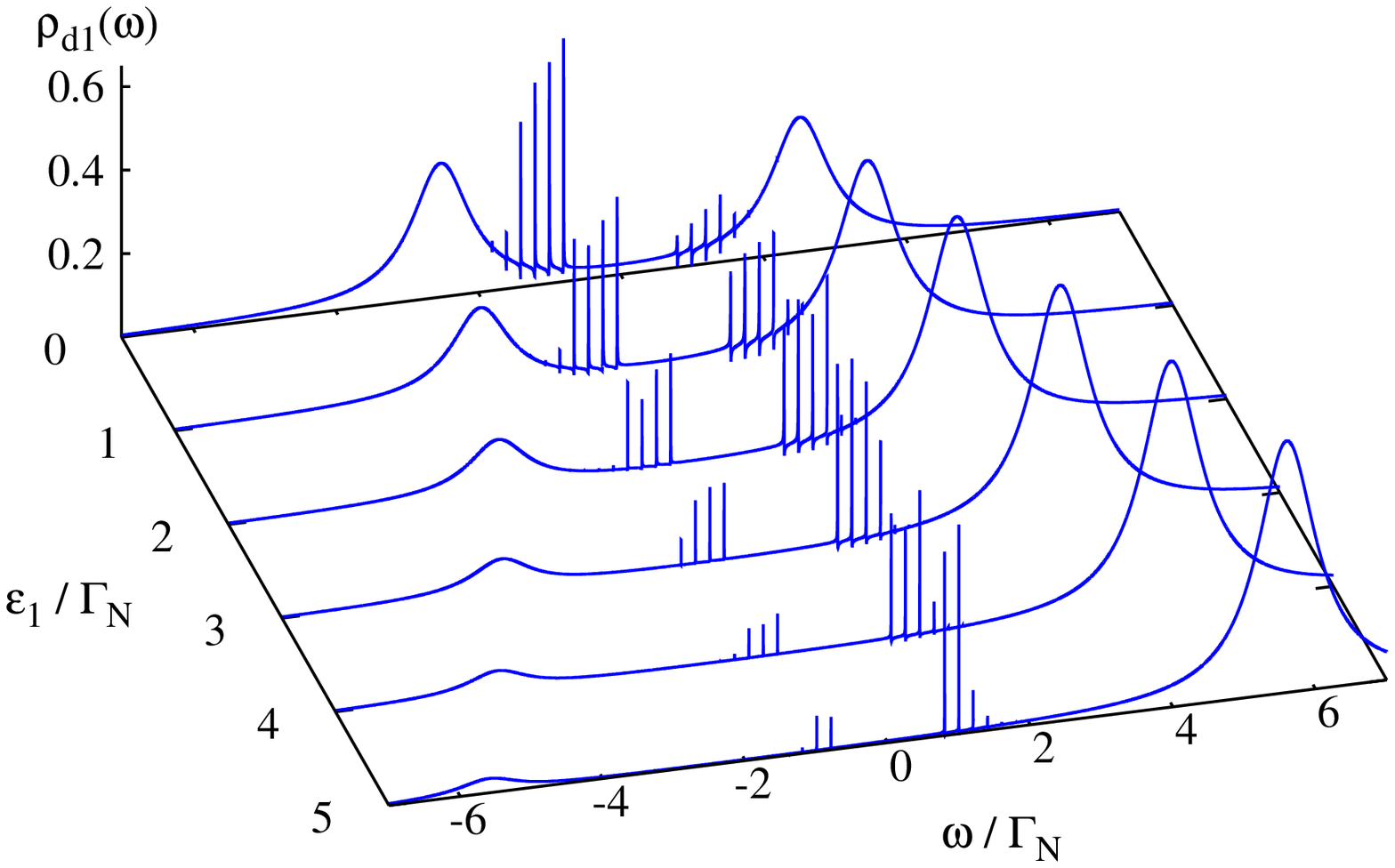}}

\vspace{-0.8cm}
\epsfxsize=9.5cm{\epsffile{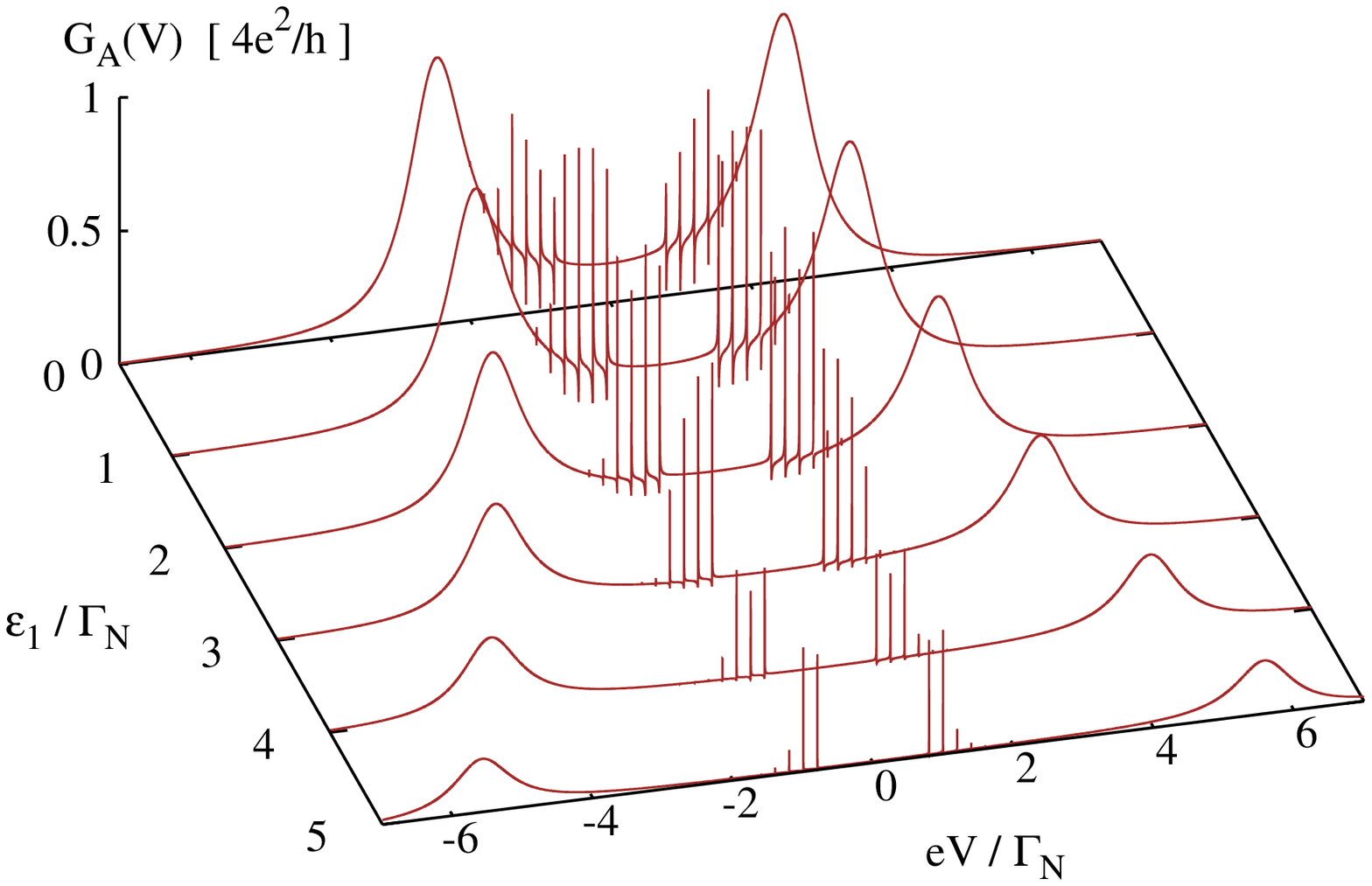}}
\caption{(color online) Dependence of the spectral function
$\rho_{d1}(\omega)$ and the Andreev conductance $G_{A}(V)$
on $\varepsilon_{1}$ (tunable by the gate voltage). Calculations
have been done for the same parameters as in figure \ref{TA_evol}.}
\label{effect_of_eps1}
\end{figure}

The subgap quasiparticle lorentzian peaks (often referred as {\em 
the bound Andreev states}) depend  on the energy level $\varepsilon_{1}$. 
In the case of single quantum dot (i.e.\ for vanishing $t$) the spectral 
function $\rho_{d1}(\omega)$ consists of two lorentzians 
at $\pm E_{1}$ (where $E_{1}= \sqrt{\varepsilon_{1}^{2}+
(\Gamma_{S}/2)^{2}}$) broadened by $\Gamma_{N}$. Their spectral weights 
are given by the BCS factors $\frac{1}{2}\left( 1 \pm 
\varepsilon_{1}/E_{1} \right)$. This fact has some importance  also
for the interferometric features. Figure \ref{effect_of_eps1} 
shows the spectrum  $\rho_{d1}(\omega)$ for several values of 
$\varepsilon_{1}$. When the energy $\varepsilon_{1}$ moves away 
from the Fermi level (by applying the gate voltage)  we observe 
a gradual redistribution of the quasiparticle spectral weights 
accompanied with suppression of the Fano resonances, especially at 
$-(\tilde{\varepsilon}_{2}+l\omega_{0})$. The Andreev transmittance 
$T_{A}(\omega)$ and differential conductance $G_{A}(V)$ are even functions 
therefore such particle-hole redistribution is not pronounced, 
nevertheless suppression of the phonon induced Fano lineshapes 
is well noticeable. 
 
\section{Correlation effects} 

In this section we address qualitative effects caused by the Coulomb
repulsion between the opposite spin electrons. Roughly speaking, we 
expect some possible signatures of the charging effect (Coulomb blockade) 
and eventual hallmarks of the Kondo physics. Since the electron transport 
occurs in our setup via QD$_{1}$ we suspect that predominantly 
the Coulomb potential $U_{1}$ can have a significant role. For completeness
we shall however study the influence of correlations on both quantum dots.

\subsection{Influence of {\boldmath $U_{2}$}}

We start by considering the effects of finite $U_{2}$, neglecting 
 correlations on the central quantum dot ($U_{1}\!=\!0$). Since the 
side-attached quantum dot is  weakly hybridized with QD$_{1}$ 
therefore the indirect influence of external leads on QD$_{2}$ 
should be rather meaningless. For this reason we impose a diagonal 
structure of the selfenergy
\begin{eqnarray} 
{\mb \Sigma}^{U}_{i}(\omega) \simeq 
\left( \begin{array}{cc}
\Sigma^{diag}_{i,\uparrow}(\omega) & 0 \\ 0 & 
-\left[ \Sigma^{diag}_{i,\downarrow}(-\omega) \right]^{*}
\end{array} \right) ,
\label{U_diagonal} 
\end{eqnarray} 
and here $i\!=\!2$. We next approximate the diagonal terms of 
(\ref{U_diagonal}) by the atomic limit solution 
\begin{eqnarray} 
\frac{1}{\omega - \varepsilon_{2} - \Sigma^{diag}_{2,\sigma}(\omega)} 
= \frac{1-n_{2,\bar{\sigma}}}{\omega\!-\!\varepsilon_{2}} 
+\frac{n_{2,\bar{\sigma}}}{\omega\!-\!\varepsilon_{2}-U_{2}} 
\label{U2_loc}
\end{eqnarray} 
where $\bar{\downarrow}=\uparrow$, $\bar{\uparrow}=\downarrow$.
It has been pointed out \cite{Cuevas-01} that the selfenergy 
defined in equation (\ref{U2_loc}) coincides with the second 
order perturbation formula  
\begin{eqnarray} 
\Sigma^{diag}_{2,\sigma}(\omega)  &=&
U_{2} \;n_{2,\bar{\sigma}} \!+\! \frac{\left( U_{2}\right)^{2}
n_{2,\bar{\sigma}}(1-n_{2,\bar{\sigma}})}{\omega-
\varepsilon_{2}-U_{2}(1-n_{2,\bar{\sigma}})}  
\label{U2_self} 
\end{eqnarray} 
and it can be generalized into more sophisticated treatments in 
the scheme of iterative perturbative theory \cite{IPT}. For 
the weak interdot coupling $t$ we expect however that corrections 
to (\ref{U2_loc},\ref{U2_self}) are not crucial. We skip here
the higher order superexchange mechanism leadind to the exotic 
Kondo effect \cite{Tanaka_exotic_Kondo} which is beyond the scope 
of our present study.

\begin{figure}
\epsfxsize=8cm\centerline{\epsffile{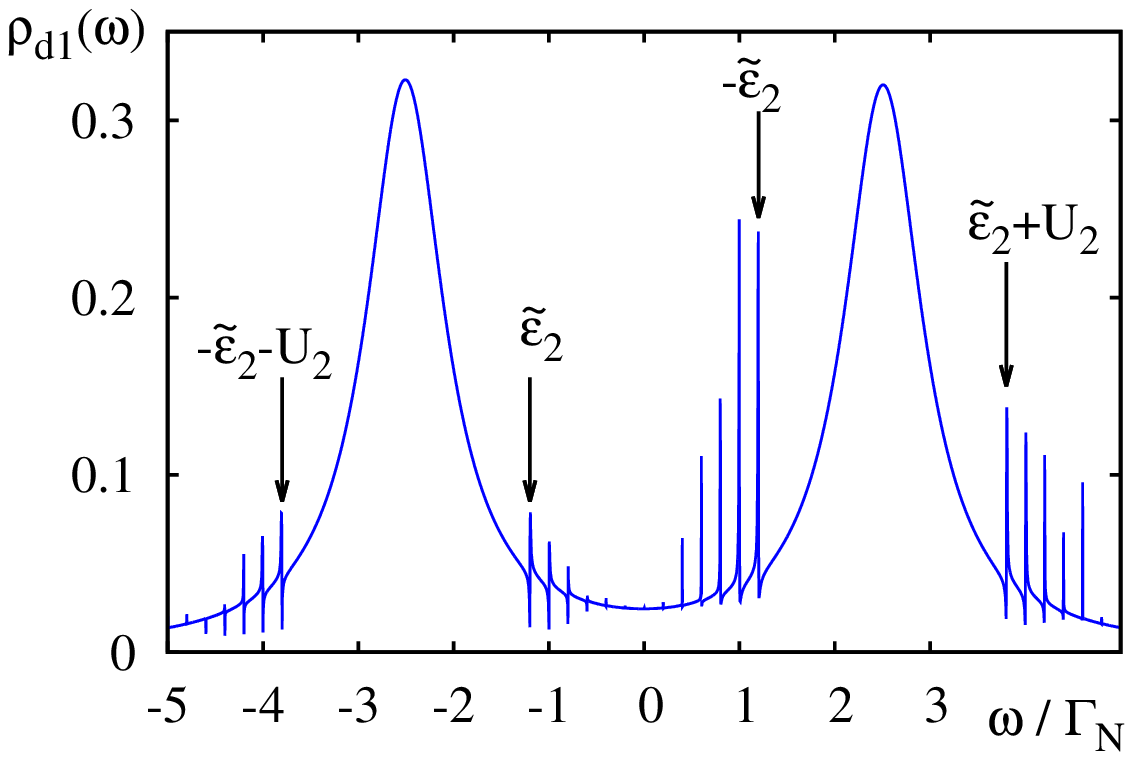}}

\vspace{-0.3cm}
\epsfxsize=8cm\centerline{\epsffile{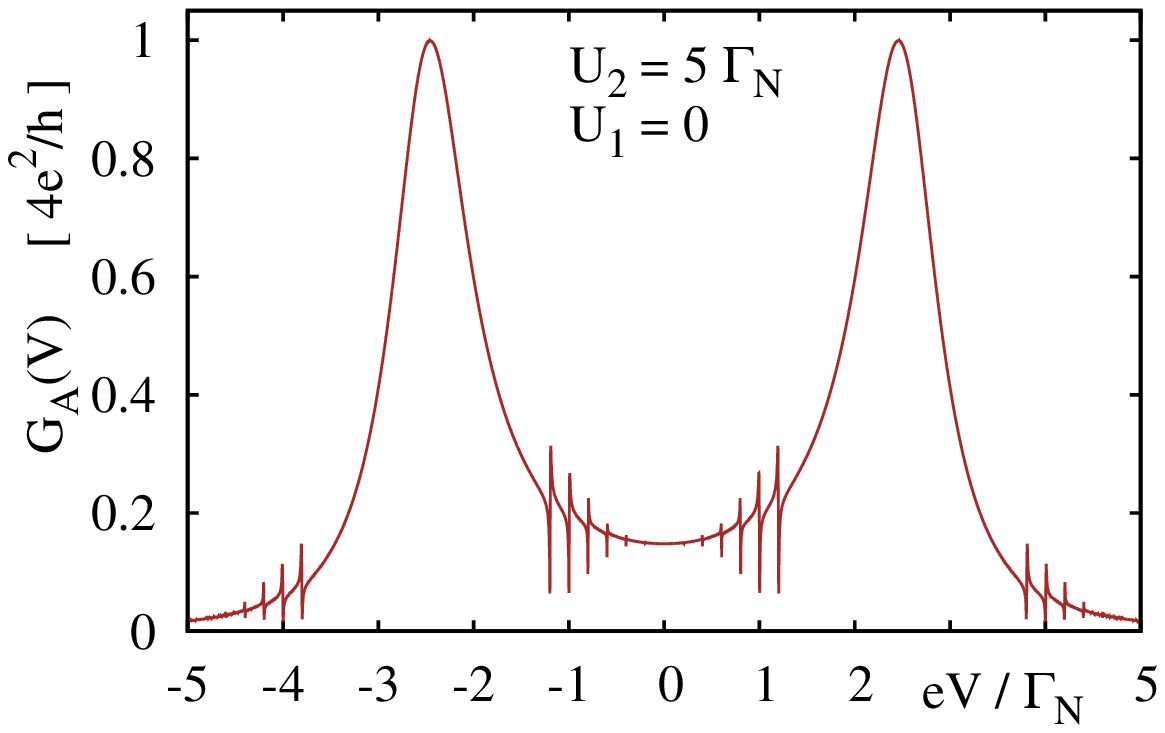}}
\caption{(color online) Spectral function $\rho_{d1}(\omega)$
of the central quantum dot (top panel) and the differential
Andreev conductance (bottom panel) obtained for $\varepsilon_{1}=0$, 
$\varepsilon_{2}=-1\Gamma_{N}$, $t=0.2\Gamma_{N}$, $g=1$,
$\omega_{0}=0.2\Gamma_{N}$, $\Delta = 10 \Gamma_{N}$
assuming the Coulomb potential $U_{2}=5\Gamma_{N}$.
The corresponding $\rho_{d2}(\omega)$ is shown in Fig.\
\ref{dos_QD2}.}
\label{dos_U2}
\end{figure}

The top panel of figure \ref{dos_U2} shows the spectrum $\rho_{d1}(\omega)$
obtained at low temperature for $U_{2}=5\Gamma_{N}$, $U_{1}=0$. As far as 
the side attached quantum dot spectrum is concerned it reveals a bunch 
of phonon peaks formed near the energy $\tilde{\varepsilon}_{2}$ 
and another group of states around the Coulomb satellite 
$\tilde{\varepsilon}_{2}\!+\!U_{2}$ (see figure \ref{dos_QD2}). These 
phonon signatures appear in $\rho_{d1}(\omega)$ as the Fano-type 
resonances. Due to the absorbed superconducting order we can notice 
effectively four groups of such Fano-type structures nearby the 
energies $\tilde{\varepsilon}_{2}$, $\tilde{\varepsilon}_{2}\!+\!U_{2}$ 
and at their mirror reflections. The Andreev transmittance $T_{A}(\omega)$ 
is symmetrized versions of what is shown in figure \ref{dos_U2} 
therefore the resulting differential conductance is  even 
function of applied voltage $V$ (see the bottom panel in Fig.\ \ref{dos_QD2}).

\subsection{Influence of {\boldmath $U_{1}$}}

Correlations originating from the Coulomb repulsion $U_{1}$ have
a totally different effect on the transport properties than above 
discussed $U_{2}$. The central quantum dot is directly coupled 
to both external leads therefore on one hand by it experiences the 
induced on-dot pairing (due to $\Gamma_{S}$) and, the other hand,  
the Kondo effect (due to $\Gamma_{N}$). These phenomena are known 
to be antagonistic. Their nontrivial competition in a context 
of the quantum dots has been discussed theoretically by many 
groups using various techniques (see Ref.\ \cite{Rodero-11} 
for a survey).

To recover basic qualitative features  we shall follow here our 
previous studies \cite{Domanski-EOM} which proved to yield satisfactory
results for the single quantum dot on interface between the metallic
and superconducting leads \cite{Deacon-10}. We choose the correlation 
selfenergy ${\mb \Sigma}_{2}^{U}(\omega)$ in the form (\ref{U_diagonal}) 
and determine its diagonal parts by the equation of motion approach 
\cite{Haug_Jauho}. Formally we use
\begin{eqnarray} 
&&\Sigma^{diag}_{1,\sigma}(\omega)  = U_{1} \left[ 
n_{1,\bar{\sigma}}\!-\!\Sigma_{1}(\omega)\right] 
\label{U1_self} \\
&&+ \frac{U_{1} \left[ n_{1,\bar{\sigma}}\!-\!
\Sigma_{1}(\omega)\right] \left[ \Sigma_{3}(\omega)
+U_{1}(1\!-\!n_{1,\bar{\sigma}})\right]}{\omega
-\varepsilon_{1}-\Sigma_{0}(\omega)-\left[ 
\Sigma_{3}(\omega)+U_{1}(1-n_{1,\bar{\sigma}})\right]} ,
\nonumber
\end{eqnarray} 
where
\begin{eqnarray}
\Sigma_{\nu}(\omega)\!\!&=&\!\!\!\sum_{{\bf k}} 
\! |V_{{\bf k} N}|^{2} \!\! \left[ \frac{1}{\omega\! - \! 
\xi_{{\bf k} N} } + \frac{1}{\omega\! -\! U_{1} \! - 
2 \varepsilon_{1}\!+\!\xi_{{\bf k} N} } \right]  
\nonumber \\
&\times & \left\{ \begin{array}{lc} 
f(\xi_{{\bf k}N}) & \mbox{\rm for } \nu=1\\
1 & \mbox{\rm for } \nu=3 \end{array}
\right. 
\label{sigmas} 
\end{eqnarray}
and as usually $\Sigma_{0}(\omega)=\sum_{{\bf k}} |V_{{\bf k} N}|^{2}
/(\omega\! - \! \xi_{{\bf k} N} )\!=\!-i\Gamma_{N}/2$.

\begin{figure}
\epsfxsize=8.5cm\centerline{\epsffile{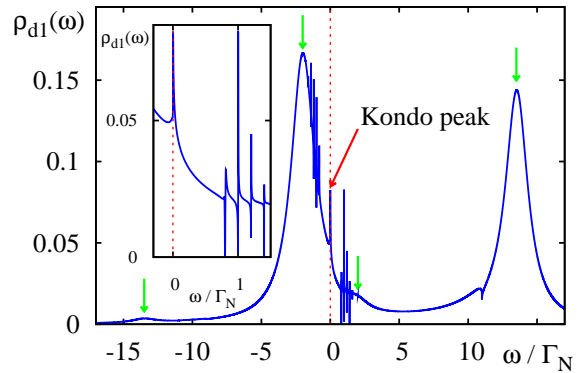}}
\caption{(color online) Spectral function $\rho_{d1}(\omega)$ 
of the central quantum dot obtained in the Kondo regime for 
$k_{B}T=10^{-3}\Gamma_{N}$, $\varepsilon_{1}=-2\Gamma_{N}$, 
$U_{2}=15\Gamma_{N}$, $\Gamma_{S}=4\Gamma_{N}$, $t=0.2\Gamma_{N}$,
$\omega_{0}=0.2\Gamma_{N}$, $g=1$, $\Delta\gg\Gamma_{N}$.
The vertical arrows indicate positions of the subgap Andreev states
at $\pm E_{1}$ and at their Coulomb satellites. We can notice that
the Kondo peak at $\omega=0$ is distinct from the
phonon induced Fano resonances.}
\label{dos_QD1_U}
\end{figure}

Let us remark that upon neglecting the terms $\Sigma_{1}(\omega)$ 
and $\Sigma_{3}(\omega)$ the selfenergy (\ref{U1_self}) nearly 
coincides with the second order perturbation formula (\ref{U2_self}) 
\begin{eqnarray} 
&&\lim_{\Sigma_{1},\Sigma_{3} \rightarrow 0} \Sigma^{diag}_{1,\sigma}
(\omega)  =\\ && U_{1}n_{1,\bar{\sigma}} \label{self_comp} +
 \frac{\left[ U_{1} \right]^{2} n_{1,\bar{\sigma}} (1\!-
\!n_{1,\bar{\sigma}})}{\omega-\varepsilon_{1}-U_{1}(1-n_{1,
\bar{\sigma}})\!-\!\Sigma_{0}(\omega)} 
\nonumber
\end{eqnarray} 
except of $\Sigma_{0}(\omega)$ present in the numerator of 
(\ref{self_comp}). This indicates that equation (\ref{U1_self}) 
is able to capture the charging effect (Coulomb blockade). The 
additional terms $\Sigma_{\nu}(\omega)$ provide corrections which 
are important in the Kondo regime, i.e.\ for  $\varepsilon_{1}
<0<\varepsilon_{1}+U_{1}$ at temperatures below $T_{K}=0.29
\sqrt{U_{1}\Gamma_{N}/2} \;\mbox{\rm exp}\left[ \frac{\pi 
\varepsilon_{1}(\varepsilon_{1}+U_{1})}{\Gamma_{N} U_{1}} 
\right]$. Under these conditions at $\mu_{N}$ there forms 
the narrow Kondo resonance of a width scaled by $k_{B}T_{K}$. 
In our present method such Kondo resonance is only qualitatively 
reproduced \cite{Domanski-EOM}. Its structure in the low energy 
regime $|\omega| \leq k_{B}T_{K}$ must be inferred from the 
renormalization group or other more sophisticated treatments.

Let us now point out the main properties characteristic 
for the Kondo regime. The equilibrium spectrum of QD$_{1}$ 
illustrated in figure \ref{dos_QD1_U} consists of four 
Andreev bound states (indicated by the vertical arrows)  
centered at $\pm \sqrt{\varepsilon_{1}^2+(\Gamma_{S}/2)^{2}}$ 
and $\pm \sqrt{(\varepsilon_{1}+U_{1})^2+(\Gamma_{S}/2)^{2}}$. 
The fact that $\varepsilon_{1}$ is located aside the superconducting 
energy gap causes asymmetry of the quasipartice spectral weights.
Besides these broad lorentzians we additionally notice the narrow peak 
at the Fermi level  signifying the Kondo effect. Such 
Kondo peak is considerably reduced in comparison to the normal case 
$\Gamma_{S}=0$ because of a competition with the on-dot pairing 
\cite{Domanski-EOM}. On top of this picture we 
recognize the phonon degrees of freedom appearing as the Fano-type 
resonances at $\pm (\tilde{\varepsilon}_{2}+l\omega_{0})$. 

\begin{figure}
\epsfxsize=9cm\centerline{\epsffile{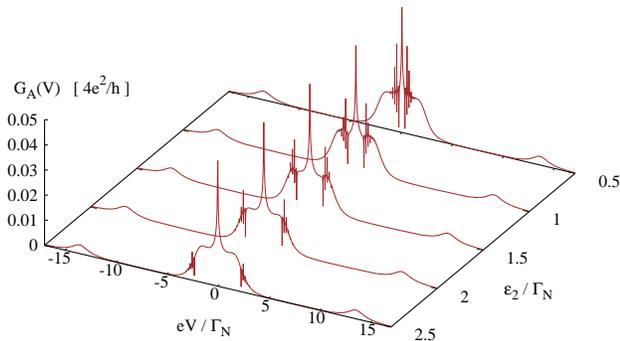}}
\caption{(color online) The differential Andreev conductance 
$G_{A}(V)$ for the set of parameters discussed in figure 
\ref{dos_QD1_U} and several values of $\varepsilon_{2}$.}
\label{cond_QD1_U}
\end{figure}

The above listed effects are also detectable in the 
differential Andreev conductance (see Fig.\ \ref{cond_QD1_U}).
$G_{A}(V)$ has local maxima at voltages  corresponding to 
the energies of the subgap bound states. Furthermore, similarly
to our previous studies \cite{Domanski-EOM}, we notice that 
the Kondo peak leads enhances the zero-bias Andreev conductance. 
This property has been indeed observed experimentally 
\cite{Deacon-10}. In the present situation we additionally
observe the Fano-type resonances. They destructively affect
the zero-bias enhancement whenever the phonon features happen 
to be located nearby the Kondo peak. The zero-bias feature
itself is also quite sensitive to the asymmetry ratio 
$\Gamma_{S}/\Gamma_{N}$ - practically it is visible only 
when both couplings $\Gamma_{\beta}$ are comparable \cite{Domanski-EOM}.

\section{Conclusions}

We have studied influence of the phonon modes on the spectral and 
on the Andreev transport  in the double quantum dot vertically
coupled between the metallic and superconducting electrodes. Our 
studies focused on the  weak interdot coupling $t$ assuming that 
phonons directly affect only the side-attached quantum dot. 
Under such circumstances an external phonon bath leads to some
interferometric effects, reminiscent of the dephasing setup 
introduced by  B\"uttiker \cite{Buttiker-88}.

In particular we find a number of the equidistant Fano-type patterns 
manifested both in the effective spectrum and in the subgap transport 
properties. These lineshapes appear at $\pm (\tilde{\varepsilon}
_{2}+l\omega_{0})$ (where $\omega_{0}$ is the phonon energy, $l$ 
is an integer number and $\tilde{\varepsilon}_{2}=\varepsilon
_{2}-\lambda^{2}/\omega_{0}$ describes the QD$_{2}$ energy shifted 
by a polaronic term). They can be regarded as {\em replicas} of the 
initial interferometric structures in absence of the phonon 
bath formed at $\pm \varepsilon_{2}$ \cite{Baranski-11,Michalek-12}. 

Electron correlations $U_{i}$ on the quantum dots can induce additional 
Coulomb satellites of these Fano features. We have investigated in 
some detail how the correlation effects get along with the Fano
interference taking into account the induced on-dot pairing. 
We notice that phonon features are sensitive to the subgap Andreev 
states (dependent on $U_{i}$) and to the Kondo effect. The latter one 
is important whenever the phonon lineshapes are induced in a vicinity 
of the Kondo peak. We thus expect that quantum interference would 
destructively affect the Kondo physics by partly suppressing its 
zero-bias hallmark \cite{Deacon-10}.

It would be of interest for the future studies to check if the presently
discussed effects are still preserved when the interdot coupling 
$t$ is comparable to the external hybridization $\Gamma_{\beta}$. 
We suspect, that the Fano-type patterns shall evolve into some new 
qualities typical for the complex molecular structures. Furthermore, 
the role of Coulomb interaction $U_{2}$ might prove to be more  
influential via the higher order exchange integrations inducing some 
exotic kinds of the indirect Kondo effect \cite{Tanaka_exotic_Kondo}.
These nontrivial issues deserve further studies eventually using 
some complementary methods.   

\begin{acknowledgments}
We acknowledge the discussions with K.I.\ Wysoki\'nski, B.R.\ Bu\l ka
and S.\ Andergassen.  The project is supported by the National Center of 
Science under the grant NN202 263138.
\end{acknowledgments}

\appendix
\section{Fano-type interference: phenomenological arguments}

Here we would like to explain in a simple way why the interferometric 
Fano structures show up in the transport properties of DQD system. In 
general, the Fano-type lineshapes \cite{Fano_exp} emerge 
whenever the localized (resonant) electron waves interfere with 
a continuum (or with sufficiently broad electron states). Following 
\cite{Fano_theor} let us consider the very instructive example in which 
the "direct" transmission channel $t_{d}=\sqrt{G_{d}}e^{i\phi_{d}}$ 
(here $G_{d}$ denotes its conductance and $\phi_{d}$ stands for 
an arbitrary phase) is combined with the transmission amplitude 
$t_{r}(\omega)=\sqrt{G_{r}}\frac{(\Gamma_{L}+\Gamma_{R})/2}
{\omega-\varepsilon_{r}+i(\Gamma_{L}+\Gamma_{R})/2}$ of another 
"resonant" level $\varepsilon_{r}$. From general considerations 
\cite{Haug_Jauho} the corresponding conductance of such "resonant" 
level is $G_{r}=\frac{2e^{2}}{h}\frac{4\Gamma_{L}\Gamma_{R}}
{\left( \Gamma_{L}+\Gamma_{R}\right)^{2}}$. Effectively these
two channels yield the following asymmetric structure
\begin{eqnarray}
G(\omega) = \left| t_{d} + t_{r}(\omega) \right|^{2} =
G_{d} \; \frac{\left| \tilde{\omega} + q \right|^{2}}
{\tilde{\omega}^{2}+1} ,
\label{Fano_shape}
\end{eqnarray}
where $\tilde{\omega}=(\omega-\varepsilon_{r})/(\frac{1}{2}\Gamma)$ 
and $q=i+e^{-i\phi_{d}}\sqrt{\frac{G_{r}}{G_{d}}}$ is the characteristic 
asymmetry factor. 

Similar reasoning can be applied to the T-shape double quantum dot 
system shown in Fig.\ \ref{scheme}. For simplicity let as neglect 
the phonon bath and assume that both electrodes are normal conductors. 
In the case of weak interdot coupling $t^{2}\ll\Gamma_{\beta}^{2}$ 
the side-attached dot QD$_{2}$ plays the role of "resonant" channel 
with its transmission amplitude $t_{r}(\omega) = \sqrt{G_{r}}
\frac{t/2}{\omega-\varepsilon_{2}\!+\!it/2}$, where $G_{r}=
\frac{2e^{2}}{h}$. On the other hand the other central dot 
QD$_{1}$ provides a relatively broad background $t_{d}(\omega) 
= \sqrt{\frac{2e^{2}}{h} \frac{4\Gamma_{N}\Gamma_{S}}{\left( 
\Gamma_{N}+\Gamma_{S}\right)^{2}}}\; \frac{(\Gamma_{N}+\Gamma_{S})/2}
{\omega-\varepsilon_{1}+i(\Gamma_{N}+\Gamma_{S})/2}$. For energies
$\omega \sim \varepsilon_{2}$ the latter amplitude is nearly constant 
$t_{d}(\omega) \simeq \sqrt{G_{d}} e^{i\phi_{d}}$ with $\sqrt{G_{d}}
=\sqrt{\frac{2e^{2}}{h} \frac{4\Gamma_{N}\Gamma_{S}}{\left( 
\Gamma_{N}+\Gamma_{S}\right)^{2}}} \left| \frac{1}{2(\varepsilon_{1}
-\varepsilon_{2})/( \Gamma_{N}+\Gamma_{S})+i} \right|$. Under such
conditions the resulting conductance $G(\omega) = \left| t_{d} 
+ t_{r}(\omega) \right|^{2}$ indeed reduces to the Fano structure 
(\ref{Fano_shape}). Some more specific microscopic arguments in 
support for the Fano-type interference of the strongly correlated 
quantum dots have been discussed at length e.g.\ by Maruyama 
\cite{Maruyama-04} and by \v{Z}itko \cite{Zitko-10}.

\end{document}